\documentclass{ws-procs9x6}

\begin{document}

\title{How Interstellar Chemistry (and Astrochemistry More Generally) Became
Useful}

\author{T. W. Hartquist$^*$ and S. Van Loo}

\address{School of Physics and Astronomy, University of Leeds,\\
Leeds LS2 9JT, United Kingdom\\
$^*$E-mail: twh@ast.leeds.ac.uk}

\author{S. A. E. G. Falle}

\address{Department of Applied Mathematics, University of Leeds,\\
Leeds LS2 9JT, United Kingdom}

\begin{abstract}
In 1986 Alex Dalgarno published a paper entitled {\it Is Interstellar
Chemistry Useful?}\cite{Da86} 
By the middle 1970s, and perhaps even earlier, Alex had
hoped that astronomical molecules would prove to: possess significant
diagnostic utility; control many of the environments in which
they exist; stimulate a wide variety of physicists and chemists who are
at least as fascinated by the mechanisms forming and removing the molecules
as by astronomy. His own research efforts have contributed greatly to
the realization of that hope. This paper contains a few examples
of: how molecules are used to diagnose large-scale dynamics in
astronomical sources including star forming regions and supernovae; the
ways in which molecular processes control the evolution of astronomical
objects such as dense cores destined to become stars and very evolved
giant stars; theoretical and laboratory investigations that elucidate the
processes producing and removing astronomical molecules and allow their
detection.
\end{abstract}


\bodymatter

\section{Introduction (by T. W. Hartquist)}
Jane Fox's eloquent comments about Alex Dalgarno's qualities as a friend,
made at the September 2008 symposium honoring him, provided a fitting
tribute to Alex's generosity, kindness and thoughtfulness. I am privileged
to have Alex as a friend and also to have benefitted from his professional
and intellectual support.

Many of Alex's former students and postdocs, as well as other colleagues,
have stories about him similar to those that I will relate. We should
all wonder how many letters of recommendation Alex has written. That
activity must have consumed a tremendous amount of time. However, Alex's
professional support of others has often gone far beyond letter writing. He
has visited many of us shortly after our moves to new positions and has
helped us make good impressions on our new colleagues. In Leeds, as
elsewhere, he served on a visiting panel providing advice on how the
local physics and astronomy research effort might be developed. The other
two panelists
visited one day each. In contrast, over a three day period, he spoke
one-on-one with every available permanent member of our Physics and
Astronomy academic staff to gain a complete and thorough overview of our
activity. He produced an insightful report, which due to its concise,
reasoned and incisive nature, carried considerable impact. It contributed
very positively to our subsequent, successful efforts to establish a new
group, with four permanent academic posts, conducting theoretical research
on fundamental quantum processes. Alex continued to help Leeds after his
2003 stay. In 2007 he served as an external member of the committee that
appointed Paola Caselli as our Professor of Astronomy.

In 1975 when I, as a postgraduate student, first worked with Alex, I soon
developed interests in hydromagnetics and plasma kinetics. Rather than
encourage me to focus only on molecular processes, Alex supported my other
interests. One problem that I wished to pursue concerns the screening of
molecular clouds from ionizing cosmic rays by scattering on Alfv\'en waves.
Alex kindly arranged for me to spend substantial fractions of the summers of
1976 and 1977 in Cambridge, England. While there in 1976, I received a
letter from him drawing my attention to a recently published paper by
Skilling and Strong\cite{Sk76} on that topic. Alex has often astounded others
with his encyclopedic knowledge of a tremendous range of literature. His
awareness of so much has often been of great help to others. Alex, Holly
Doyle and I made use of results in
Ref.~\refcite{Sk76} in a paper on
ionization rates inferred for diffuse molecular clouds.\cite{Ha78b}

The title of the present paper echoes that of a paper by Alex\cite{Da86} 
to which John Black and Ewine van Dishoeck also referred during their talks at
the September 2008 symposium. The remainder of this paper is divided into
six sections, half of which mention selected contributions of Alex primarily
in molecular astrophysics. My coauthors and I have divided his contributions
into those
concerning molecular diagnosis, those showing how molecular processes
control astrophysical environments and those in the investigation of
quantum processes relevant to astrophysics. Sections addressing selected
recent works on these themes by other researchers interleave with those
summarizing Alex's efforts. I apologize to Alex and to others for the fact
that the selections cannot be comprehensive. We have had to neglect a great
deal of excellent work done by Alex and by others. Ewine gave a talk at the
symposium in which she summarized some of Alex's work in astrochemistry and
some recent work of others. The overlaps and the differences between what
she said and what I said demonstrate the strength and breadth of Alex's work
and of the current state of molecular astrophysics.

\section{Alex's Work on Diagnostics}
The Copernicus satellite, launched in August 1972, enabled far ultraviolet
spectroscopy of the nearest O and B stars and of diffuse interstellar matter
along the lines of sight to them. Alex and his students and other
collaborators developed the framework and tools to use Copernicus data to
probe the natures of diffuse molecular clouds, those molecular clouds
having optical depths at visual wavelengths less than about unity. Black and
Dalgarno\cite{Bl77} 
interpreted absorption measurements of the column densities
of atomic hydrogen, the ground and next lowest six rotational levels of
H$_2$, HD and OH for the line of sight to $\zeta$ Oph. They inferred
the intensity of the far ultraviolet radiation impinging on the intervening
cloud, the thermal and density structure, the elemental deuterium abundance
and the rate at which cosmic rays induce ionization in the cloud. A large
body of work on the relevant quantum processes, begun in the late 1960s by
Alex and collaborators, underpins the construction of models like that of
Black and Dalgarno\cite{Bl77}.
Some of the work on processes is described later.
Alex had to possess a profound long term vision to construct the foundations
upon which the models were built and then to develop and apply the models.

Knowledge of the elemental deuterium abundances in different regions
constrains cosmological models and provides insight into the effect of 
astration on elemental abundances. Studies of the spatial variation of the
cosmic ray induced ionization rate are relevant to the understanding of 
the acceleration and propagation of cosmic rays. Some of Alex's additional
findings on deuterium and cosmic rays in diffuse clouds are described in 
papers by Black and Dalgarno\cite{Bl73}
and Hartquist, Black and Dalgarno\cite{Ha78a}.

The abundance ratios of some deuterated species, including DCO$^+$, relative
to their protonated counterparts, serve as diagnostics of the fractional
ionization in dark molecular clouds. As described more fully later, 
the fractional ionization governs the role of the magnetic field in star
formation in a dark cloud. Dalgarno and Lepp\cite{Da84}
showed that in dark
clouds sufficient deuterium is in atomic form to affect the deuterium
fractionation. The inference of reliable constraints on fractional
ionizations in dark clouds from observations of deuterated and protonated
isomers requires consideration of key reactions involving atomic deuterium.

The interaction of cosmic rays with H$_2$ in interstellar clouds deposits
energy and induces excitation. In the Jovian atmosphere, the energy 
deposition and excitation due to particles accelerated in the solar wind -
magnetosphere interaction results in observable H$_2$ ultraviolet emission.
Alex's work on the interpretation of Jovian auroral and airglow emissions
\cite{Li96a,Li96b}
demonstrated the presence of substantial
temperature gradients. We refrain from mentioning other Solar System related
studies conducted by Alex, as others writing articles for this volume are
addressing those topics. However, a mention of the Jovian work here is
appropriate because the approach Alex took in it bears a relationship to that
he adopted in some investigations of deposition in extra-Solar System
objects, including SN1987A.

Monitoring of the SN1987A supernova ejecta revealed the presence of CO
infrared line emission at 112 days. Liu, Dalgarno and Lepp\cite{Li92} 
developed
a model of the thermal balance and chemistry of the ejecta. Heating is due
to radioactive decays which produce $\gamma$-rays, and their
interaction with matter generates energetic electrons. 
Radiative association forms CO. The
very significant conclusion drawn about the ejecta's
dynamics is that microscopic mixing of helium-rich layers with layers rich
in carbon and oxygen had to be at most very limited, despite the full
development of the Rayleigh-Taylor instability during the explosion.
Otherwise, microscopic mixing would have led to the destruction of CO by
He$^+$ at a rate incompatible with the observations.

\section{Others' Work on Diagnostics}
Studies of cosmic ray ionization rates continue. Observations of H$_3^+$
infrared absorption now provide constraints on the rates which imply they
are an order of magnitude higher in some diffuse interstellar clouds than
inferred by Alex and his collaborators in the 1970s.\cite{In07} 
The differences are due to the adoption in the 1970s of a value for the
H$_3^+$ dissociative recombination rate coefficient that is small compared
to that now accepted. (See section 7 for a mention of the relevant
experimental measurements.) The analysis by Caselli $et$ $al.$\cite{Ca98b,Ca02}
of millimeter observations of deuterated species and their protonated
counterparts has advanced our knowledge of the variation of the fractional
ionization in dense molecular cores which are evolving to form stars.

The inference of the dynamics of dense core collapse from millimeter and
submillimeter molecular line observations is a very challenging problem.
Given that the relative roles of hydromagnetic wave processes and gravity at
different stages of collapse continue to be debated, that problem is
important. Its solution requires the construction of appropriate
dynamical models, the development of accurate models for the gas phase
and surface chemistries of dense cores, and involved radiative transfer
calculations. Data for multiple lines of multiple species are required.
For a simple dynamical description, Tsamis $et$ $al.$\cite{Ts08}
have performed
relevant chemical and radiative transfer calculations. Much work remains,
but their
study gives a good indication of what is required.

The angular resolution and sensitivity of ALMA will enable the mapping of
molecular distributions in protoplanetary discs. The study of the chemistry
of the discs is at an early stage, and a full understanding of what can be
learned about disc dynamics from molecular observations does not exist.
Ilgner $et$ $al.$\cite{Il04}
are amongst those who have begun to address this issue
by considering chemistry for an $\alpha$ disc model. The use of more
complicated models of disc dynamics, including gravitational instability and
the effects that the magnetic rotational instability has on viscosity, would
be interesting. Thi, van Zadelhoff and van Dishoeck\cite{Th04}
performed a notable observational study of simple organic molecules in
protoplanetary discs around T-Tauri and Herbig Ae stars with single dish
millimeter and submillimeter telescopes. Lahuis $et$ $al.$\cite{La06}
detected
emission from hot organic molecules in a protoplanetary disc with the
Spitzer Space Telescope. From data obtained for the disc of TW Hydrae with
the Submillimeter Array, Qi $et$ $al.$\cite{Qi08}
have concluded that disc chemical
models should include active deuterium fractionation. Existing results
point to an interesting future for the diagnosis of protoplanetary discs
with molecular observations.

\section{Alex's Work on Chemical and Quantum Control}
Though this article primarily concerns molecular astrophysics, some of
Alex's contributions to astrophysics that did not address only molecules
are too significant to exclude. The Dalgarno and McCray\cite{Da72}
paper is
known to astrophysicists working on a wide variety of nonmolecular sources
including supernova remnants. Its title {\it Heating and Ionization of HI
Regions} is a bit deceptive. HI regions are the main topic of the paper,
but the cooling of gas at temperatures far higher than those of HI regions
is also addressed in it. Its famous Figure 2 shows the cooling rate
coefficient for gas from temperatures of about 20K up to $10^8$K. Dalgarno
and McCray\cite{Da72} 
considered molecular coolants but atomic processes received
more space than molecular processes. They also summarized research initiated
by themselves and C. Bottcher and M. Jura. Bottcher $et$ $al.$\cite{Bo70}
showed
that supernovae and temporal variations in the population of hot stars are
frequent enough that they prevent the interstellar medium from reaching a
state in which gas is in two coexisting phases in pressure, ionization and
thermal equilibrium. Later in the 1970s, the role of supernovae in
establishing the global properties of the interstellar medium became a major
topic in interstellar research. Alex and his collaborators pioneered the
investigation of this important aspect of galactic astronomy.

Alex's work on heating and cooling has extended to the molecular cooling of
the Early Universe. Stancil, Lepp and Dalgarno\cite{St98}
examined the deuterium
chemistry of the pregalactic and protogalactic eras. Though much less
abundant than H$_2$, HD was possibly an important coolant then, due in
part to its possession of a dipole moment and H$_2$'s lack of one. This
difference between HD and H$_2$, and the greater mass that D has than
H has, leads to the lowest excited level of HD that can be populated
by collisions with H having a roughly four times lower energy than the
lowest excited level of H$_2$ that can be populated by collisions with H.
Thus, as the temperature drops, HD becomes an increasingly more effective
coolant than H$_2$. HD's dipole moment causes the radiative decay
rates of its excited levels to be much larger than those of the
corresponding H$_2$ levels. Consequently, the cooling rate per HD molecule
continues to increase with increasing H number density to a higher
density than the cooling rate per H$_2$ molecule does.

Lepp and Dalgarno\cite{Le88}
performed early work on the role that photoabsorption
by large molecules or large molecular ions (e. g. free flying polycyclic
aromatic hydrocarbons and their negative ions) play in heating interstellar
clouds. At the time of their work grain photoelectric heating had been
advocated as the primary heating mechanism in diffuse clouds, but it was
appearing to be insufficient in at least some sources. Lepp and Dalgarno\cite{Le88} 
identified photodetachment from large negative molecular ions as an
important previously unconsidered heating process.

Section 2 contains a mention of the relevance of the fractional ionization to
the part that magnetic fields have in molecular cloud dynamical evolution and
star formation. Oppenheimer and Dalgarno\cite{Op74}
developed a model of the
chemistry governing the fractional ionisation in dark molecular regions. One
of their key realizations is that charge transfer of molecular ions with
neutral metal atoms causes a significant reduction in the gas phase
recombination rate. 

Star forming regions contain shocks driven by the winds and jets of young
stars. In many cases these shocks may greatly influence subsequent star
formation. In the second half of the 1970s and during the 1980s, the study
of the shocks through the observation of infrared H$_2$ line emission at
about 2 microns and of millimeter line emission from other molecules became
a major industry. The theoretical models of the shocks had to
increase in sophistication to match observational progress. Multifluid
hydromagnetic models were developed. Draine, Roberge and Dalgarno\cite{Dr83}
contributed significantly in this area by constructing such models and
by critically evaluating and compiling data for the processes controlling
the thermal balance. The chemistry controlling the ionization balance has
a major impact on the shock structure, because the size of the dissipation
region depends on the number density of ions. In the paper by Pineau des
Forets $et$ $al.$\cite{Pi86}
Alex identified the chemistry controlling the
ionization structure of shocks in diffuse clouds.

Alex's studies of how chemistry controls the thermal balance of
astrophysical sources has included the work in which he explained the
existence of large inhomogeneities in the thermal structure of the
SN1987A ejecta as the consequence of the variation of the chemical structure
throughout the oxygen core.\cite{Li95}

\section{Others' Work on Chemical and Quantum Control}
During the 1980s and 1990s T. Ch. Mouschovias and his collaborators, F. H.
Shu and his collaborators and others developed a picture of the birth
of solar-mass stars in dense molecular cores undergoing gravitationally
induced, quasistatic collapse regulated by the magnetic field and ambipolar
diffusion. Ambipolar diffusion is the motion of charged species relative to
neutrals driven by the magnetic force. In this picture the core
forms with a magnetic flux to mass ratio that is too large for
gravitationally induced collapse to proceed to higher densities if the flux
to mass ratio does not alter. The core is said to be magnetically
subcritical. However, due to the low fractional ionization, ambipolar
diffusion occurs and reduces the magnetic flux, allowing low Alfv\'enic Mach
number collapse. The ambipolar diffusion timescale depends on the number
density of ions, which is the prime reason for the earlier mentions in this
article of the fractional ionization in dense clouds.

In the late 1990s, the picture described in the previous paragraph was
challenged. For instance, Myers and Lazarian\cite{My98}
introduced a "turbulent
cooling flow" description of the formation and evolution of dense cores.
They wished to account for the observed linewidths of features arising in the
envelopes of cores being broader than expected from some magnetic and
ambipolar diffusion regulated gravitationally induced collapse models.

Ward-Thompson $et$ $al.$\cite{WT07}
have summarized the issues in the debate about
the roles of waves and of ambipolar diffusion operating with gravity in the
birth of solar-mass stars. (We shall refer to waves rather than turbulence,
which is a misnomer because the energy input scale and dissipation scale do
not differ sufficiently for turbulence to develop fully.) Good evidence for
wave induced evolution exists. However, good evidence for phases during
which ambipolar diffusion and gravity operate together also exists
(e. g. Chiang $et$ $al.$\cite{Ch08}).

The simulations of Tassis and Mouschovias\cite{Ta05}
are representative of work
by the Illinois group on dynamics showing the important role that the
fractional ionization and charged grains play in the gravitationally induced,
ambipolar diffusion regulated collapse phase. The simulations are of thin discs
in which the structures vertical to the symmetry planes are in magnetostatic
equilibrium.

Van Loo, Falle and Hartquist\cite{VL08}
have demonstrated the importance of
ambipolar diffusion and, hence, of the chemistry controlling the fractional
ionization during the initial formation of dense cores by the nonlinear
steepening of hydromagnetic waves. The relative roles of wave processes
and gravitationally induced, magnetically regulated collapse at different
phases of core evolution may be worthy of debate. In contrast, the fact that
the chemistry controlling the fractional ionization is of great importance
in core evolution is very clear.

\begin{figure}
\psfig{file=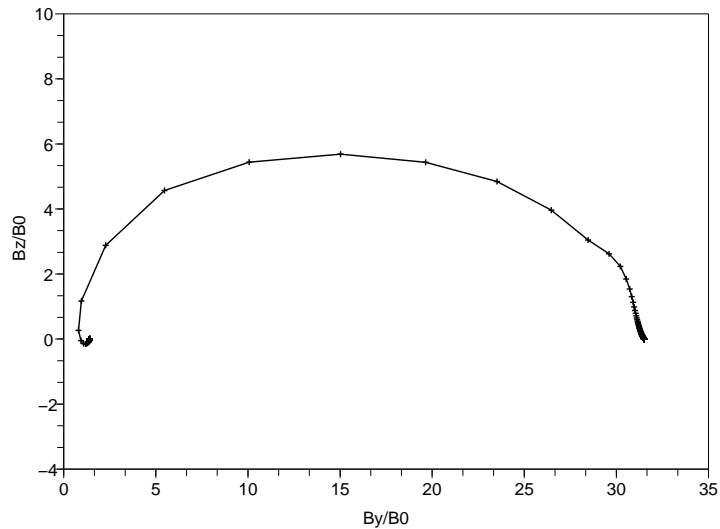, width= 4.5in}
\caption{The rotation of the magnetic field within the shock front of 
C-type shock. The shock is propagating through a quiescent medium of 
$n_{\rm H} = 10^6\ {\rm cm^{-3}}$ and $B_0 = 10^{-3}$G at 25~km\ s$^{-1}$. 
The angle between the shock propagation and the upstream magnetic field is
$45^\circ$. The crosses represent the grid spacing of the numerical model.}
\label{fig:model}
\end{figure}

Multifluid models of shocks in dark regions having hydrogen number densities
exceeding about $10^5$ cm$^{-3}$ advanced in complexity during the 1990s, and
only now are we on the verge of constructing reliable models of such shocks.
The very important Draine $et$ $al.$\cite{Dr83} 
work concerned shocks
that are propagating perpendicularly to the magnetic fields. Of
course, real shocks propagate obliquely to the magnetic fields. In dense
regions, the collisions of neutrals with charged grains rather than the
collisions of neutrals with ions dominate the coupling of neutral flows
to magnetic fields. Draine $et$ $al.$\cite{Dr83}
treated the effects due to charged
grains in an approximate fashion that is suitable for a wide variety of 
parameters but is not reliable for regions with number densities above
above about $10^5$ cm$^{-3}$. Pilipp and Hartquist\cite{Pi94} 
and Wardle\cite{Wa98}
initiated efforts to include charged grains more rigorously in models of
oblique shocks.

Wardle\cite{Wa98} showed that an attempt to solve the time-independent
coupled ordinary differential equations governing steady plane-parallel 
shock structures through the integration in a downstream direction
from upstream boundary conditions cannot yield solutions for fast-mode
shocks. Rather integration has to proceed in the upstream direction from 
downstream conditions. This is not possible unless equilibrium conditions
obtain at all points in the flow. Unfortunately, they do not. For instance,
the abundance of H$_2$O, which is an important coolant, does not attain its
equilibrium value as gas cools from several hundred degrees until it has
been at about 10K for of the order of $10^5$ years or more. This is
much longer than the flow time through a shock in a dark dense core.
Consequently, a time-dependent approach is necessary. Such an approach
including all electric current components in a proper fashion\cite{Fa03} 
has been combined only recently with a self-consistent treatment of the
chemistry controlling the fractional ionization and grain charges and,
hence, the coupling between neutral flow and the magnetic field.\cite{VL09}
Figure~\ref{fig:model} shows the rotation of the magnetic
field within the shock front of a steady C-type shock. The spiral node 
upstream is one of the reasons why Wardle\cite{Wa98}
needed 
to integrate the steady-state equations in an upstream direction to 
find a steady fast-mode shock. 
 
Chemistry plays a huge role in controlling the mass loss from highly evolved
stars. Ferrarotti and Gail\cite{Fe06}
conducted a comprehensive theoretical
study of the formation of dust in Asymptotic Giant Branch (AGB) stars.
Radiation pressure on the dust greatly influences the mass loss rates and
terminal speeds of the outflows. Mass loss rates affect the stellar
evolution and the nucleosynthesis products.

\section{Alex's Work on Quantum Processes Relevant to Astrophysics}
Alex's contributions to the study of processes important for the diagnosis
and evolution of astrophysical sources are vast.

The volume of just his work
on radiative processes important for modelling diffuse clouds is impressive.
This
includes the calculation of radiative probabilities for the Lyman
and Werner bands of H$_2$.\cite{Al69,Da70}
These are important for photodissociation and the pumping of excited
rovibrational levels of ground electronic state of H$_2$. Turner,
Kirby-Docken and Dalgarno\cite{Tu76} computed H$_2$ ground electronic state
rovibrational transition probabilities that help determine the
level populations resulting from the cascade following pumping. They are
also important for radiative cooling in shocked molecular gas in star
forming regions. Other work on radiative mechanisms important in diffuse
clouds includes that of van Dishoeck, van Hemert and Dalgarno\cite{vD84}
who calculated the OH photodissociation rate.

Collisional processes are also important for H$_2$ level populations in
diffuse clouds. Dalgarno, Black and Weisheit\cite{Da73} identified the role of
H$_2$ collisions with H$^+$ in establishing the H$_2$ ortho-para ratio in
diffuse clouds. They also identified H$_2$ collisions with D$^+$ as the
dominant mechanism for forming HD in diffuse clouds.

The work of Chu and Dalgarno\cite{Ch75} on collisional excitation of CO
is relevant to the cooling and observational diagnosis of most detectable
diffuse molecular material in the Universe. The adoption by
Roberge and Dalgarno\cite{Ro82} of a master
equation approach for the calculation of the populations of rovibrational
levels of H$_2$ in shocked gas led to a major advance in the understanding
of the collisional dissociation of astrophysical molecules.

Alex has contributed many important studies of charge transfer processes
of which that by Butler, Heil and Dalgarno\cite{Bu80} is a good example. The
efforts of this trio at the end of the 1970s and the start of the 1980s
had a huge impact on models of photoionized regions from planetary nebulae
to the broad emission line regions of quasistellar objects. Their work
led to the understanding of how charge transfer greatly
influences the emission spectra of such plasmas. Of course, Alex's
more recent investigations, described elsewhere in this volume, of charge
transfer have had major significance for research on the heliosphere,
planetary atmospheres, comets and the interstellar soft X-ray background.

Supernova ejecta belong to the long list of different classes of
astrophysical sources for which his investigation of quantum processes
are relevant. For instance, Dalgarno, Du and You\cite{Da90} calculated
the rates of the radiative association of O with C and C$^+$ as part
of Alex's program on molecules in the ejecta of SN1987A.

\section{Others' Work on Quantum Processes Relevant to Astrophysics} 

The advance of molecular astrophysics has relied on many laboratory efforts
involving a wide variety of techniques to obtain reliable reaction rate
coefficients for temperatures ranging from about 10K to over 10$^3$K.
For over two decades CRESU devices, in which expansion in supersonic jets
cools gas, have facilitated low temperature measurements relevant for the
chemistry of star forming regions. Though presented in the context of
work on Titan's atmosphere, the results obtained by Berteloite $et$ $al.$
\cite{Be08} 
for C$_4$H reactions with hydrocarbons provide recent examples of the type
of data relevant for star forming regions that can be obtained with CRESU
devices.

Storage ring experiments have produced a smaller volume of data relevant to
astrophysics, but in some cases the storage ring results have had
considerable impact. The measurement of the H$_3^+$ dissociative
recombination rate by McCall $et$ $al.$\cite{Mc03} has been important for the
inference of cosmic ray ionization rates in diffuse clouds from observations
of H$_3^+$, which we mentioned earlier.

The original masthead of {\it The Astrophysical Journal} described it as
{\it An International Review of Spectroscopy and Astronomical Physics}.
Spectroscopy plays a key role in molecular astrophysics just as it does
in other areas of astrophysics. An excellent example of the importance
of laboratory molecular spectroscopy for astrophysics is provided by the
work done by McCarthy $et$ $al.$\cite{Mc06} which led to the first detection
of an astrophysical negative molecular ion. C$_6$H$^-$ exists in the
stellar envelope of IRC +10216 and in the dense molecular cloud TMC-1.

Theoretical calculations of line wavelengths and radiative transition rates
for H$_2$O\cite{Ba06}
made possible the first detection of water
in the atmosphere of an extrasolar planet.\cite{Ti07} 
Theoretical line list calculations (e. g. Harris $et$ $al.$\cite{Ha08}) are
important for work on the atmospheres of cool stars and low
metallicity stars and distinguishing brown dwarves from planets.

To this point we have mentioned only gas phase processes. In the past
decade the effort to understand grain surface processes affecting
astrophysical chemistry has intensified. A key problem concerning the
kinetics of interstellar grain surface reactions arises because the number
of reactant atoms and molecules on a surface is often too small for a
standard rate equation treatment to be appropriate. Caselli, Hasegawa
and Herbst\cite{Ca98a}
introduced a modified rate equation approach, which they have
subsequently improved. Their 1998 paper triggered a considerable amount
of research on how to treat interstellar grain surface chemistry. Green
$et$ $al.$\cite{Gr01} adopted a master equation approach with which they
calculated the probabilities that a grain contains specific numbers of atoms
and molecules of the different reactant species. The method is suitable
for some simple problems but is too computationally expensive for many
of interest. Barzel and Biham\cite{Ba07} have developed a technique based on the
use of equations they obtained by taking moments of the master equation.
It shows considerable promise because it has proven accurate in test
calculations and computationally tractable for problems involving many
reactant species.

Laboratory studies of surface processes important for astrochemistry have
included investigations of H$_2$ formation and desorption of ices. The
short but informative review by Williams $et$ $al.$\cite{Wi07} provides a good
introduction to some of the research. While they made a good attempt to
note work done elsewhere, the main focus of the Williams $et$ $al.$\cite{Wi07} 
review is research carried out by several UK groups. Several excellent
surface chemistry groups exist throughout the world. The Leiden group is
one. Recent work in that group has included a study of the photodesorption
of CO ice.\cite{Ob07}

The formation of astrophysical dust in carbon-rich stellar envelopes is
better understood than its production in oxygen-rich envelopes.
Uncertainties in the field are sufficient that Nuth and Ferguson\cite{Nu06}
felt moved to proclaim, in the title of a paper on their relevant
experimental results, that {\it Silicates Do Nucleate in Oxygen-Rich
Circumstellar Outflows ...} Astronomical observations had already
established this fact, but the problem of explaining it has
been so challenging that Nuth and Ferguson\cite{Nu06} had some grounds for
announcing their progress in the manner they did. Theoretical efforts to
identify the species important in the nucleation of astrophysical dust
involve the development and application of methods to evaluate the energies
of multiple configurations of various numbers of particles of the nucleating
species. Bhatt and Ford\cite{Bh07} have recently published theoretical results
on the investigation of MgO as a possible nucleating species around M stars.

\section{Closing Remarks}

The unusually lengthy list of keywords at the start of this article
reflects the broad range of the work that Alex has done in astronomy
and the breadth of its impact.

One does not need to possess great prognostic powers to know that molecular
astrophysics has a bright future. Herschel is due to be launched in
2009 and ALMA will be operational well before Alex turns 90. When
we meet to celebrate that occasion, we will have learned a lot more
about astrophysical molecular sources through the use of those two
facilities.


\begin{thebibliography}{9}
\bibitem{Da86} A. Dalgarno, {\em QJRAS}, {\bf 27}, 83 (1986).
\bibitem{Sk76} J. Skilling and A. W. Strong, {\em A\&A}, {\bf 53}, 253
(1976).
\bibitem{Ha78b} T. W. Hartquist, H. T. Doyle and A. Dalgarno, {\em A\&A},
{\bf 68}, 65 (1978).
\bibitem{Bl77} J. H. Black and A. Dalgarno, {\em ApJ Suppl}, {\bf34}, 405
(1977).   
\bibitem{Bl73} J. H. Black and A. Dalgarno, {\em ApJ}, {\bf 184}, L101
(1973).
\bibitem{Ha78a} T. W. Hartquist, J. H. Black and A. Dalgarno,
{\em MNRAS}, {\bf 185}, 643 (1978).
\bibitem{Da84} A. Dalgarno and S. Lepp, {\em ApJ}, {\bf 287}, L47 (1984).
\bibitem{Li96a} W. Liu and A. Dalgarno, {\em ApJ}, {\bf 462}, 502 (1996a).
\bibitem{Li96b} W. Liu and A. Dalgarno, {\em ApJ}, {\bf 467}, 446 (1996b).
\bibitem{Li92} W. Liu, A. Dalgarno and S. Lepp, {\em ApJ}, {\bf 396},
679 (1992).
\bibitem{In07} N. Indriolo, T. R. Geballe, T. Oka and B. J. McCall,
{\em ApJ}, {\bf 671}, 1736 (2007).
\bibitem{Ca98b} P. Caselli, C. M. Walmsley, R. Terzieva and E. Herbst,
{\em ApJ}, {\bf 499}, 234 (1998).
\bibitem{Ca02} P. Caselli, C. M. Walmsley, A. Zucconi, M. Taffalla,
L. Dore and P. C. Myers, {\em ApJ}, {\bf 565}, 344 (2002).
\bibitem{Ts08} Y. G. Tsamis, J. M. C. Rawlings, J. A. Yates and S. Viti,
{\em MNRAS}, {\bf 388}, 898 (2008).
\bibitem{Il04} M. Ilgner, Th. Henning, A. J. Markwick and T. J. Millar,
{\em A\&A}, {\bf 415}, 643 (2004).
\bibitem{Th04} W.-F. Thi, G.-J. Zadelhoff and E. F. van Dishoeck, {\em A\&A},
{\bf 425}, 955 (2004).
\bibitem{La06} F. Lahuis, E. F. van Dishoeck, A. C. A. Boogert, K. M.
Pontoppidan, G. A. Blake, C. P. Dullemond, N. J. Evans II, M. R.
Hogerheijde, J. K. Jorgensen, J. E. Kessler-Silacci and C. Knez,
{\em ApJ}, {\bf 636}, L145 (2006).
\bibitem{Qi08} C. Qi, D. J. Wilner, Y. Aikawa, G. A. Blake and M. R.
Hogerheijde, {\em ApJ}, {\bf 681}, 1396 (2008).
\bibitem{Da72} A. Dalgarno and R. A. McCray, {\em ARA\&A}, {\bf 10},
375 (1972).
\bibitem{Bo70} C. Bottcher, R. A. McCray, M. Jura and A. Dalgarno, {\em ApL },
{\bf 6}, 237 (1970).
\bibitem{St98} P. C. Stancil, S. Lepp and A. Dalgarno, {\em ApJ}, {\bf 509},
1 (1998).
\bibitem{Le88} S. Lepp and A. Dalgarno, {\em ApJ}, {\bf 335}, 769 (1988).
\bibitem{Op74} M. Oppenheimer and A. Dalgarno, {\em ApJ}, {\bf 192}, 29
(1974).
\bibitem{Dr83} B. T. Draine, W. G. Roberge and A. Dalgarno, {\em ApJ},
{\bf 264}, 485 (1983).
\bibitem{Pi86} G. Pineau des Forets, D. R. Flower, T. W. Hartquist and A.
Dalgarno, {\em MNRAS}, {\bf 220}, 801 (1986).
\bibitem{Li95} W. Liu and A. Dalgarno, {\em ApJ}, {\bf 454}, 472 (1995).
\bibitem{My98} P. C. Myers and A. Lazarian, {\em ApJ}, {\bf 507}, L157
(1998).
\bibitem{WT07} D. Ward-Thompson, P. Andre, R. Crutcher, D. Johnstone,
T. Onishi and C. Wilson, {\it An Observational Perspective of Low-Mass Dense
Cores II: Evolution toward the Initial Mass Function}, in {\it Protostars
and Protoplanets V}, eds. B. Reiputh, D. Jewitt and K. Keil, (University
of Arizona Press, 2007), pp. 33-46. 
\bibitem{Ch08} H.-F. Chiang, L. W. Looney, K. Tassis, L. G. Mundy and T. Ch.
Mouschovias, {\em ApJ }, {\bf 680}, 474 (2008).
\bibitem{Ta05} K. Tassis and T. Ch. Mouschovias, {\em ApJ}, {\bf 618},
769 (2005).
\bibitem{VL08} S. Van Loo, S. A. E. G. Falle, T. W. Hartquist and A. J.
Barker, {\em A\&A}, {\bf 484}, 275 (2008).
\bibitem{Pi94} W. Pilipp and T. W. Hartquist, {\em MNRAS}, {\bf 267}, 801
(1994).
\bibitem{Wa98} M. Wardle, {\em MNRAS}, {\bf 298}, 507 (1998).
\bibitem{Fa03} S. A. E. G. Falle, {\em MNRAS}, {\bf 344}, 1210 (2003).
\bibitem{VL09} S. Van Loo, I. Ashmore, P. Caselli, S. A. E. G. Falle and
T. W. Hartquist, {\em MNRAS}, submitted (2009).
\bibitem{Fe06} A. S. Ferrarotti and H.-P. Gail, {\em A\&A}, {\bf 447},
553 (2006).
\bibitem{Al69} A. C. Allison and A. Dalgarno, {\em JQSRT}, {\bf 9},
1543 (1969).
\bibitem{Da70} A. Dalgarno and T. L. Stephens, {\em ApJ}, {\bf 160},
L107 (1970).
\bibitem{Tu76} J. Turner, K. Kirby-Docken and A. Dalgarno, {\em ApJ Suppl},
{\bf 35}, 281 (1976).
\bibitem{vD84} E. F. van Dishoeck, M. C. van Hemert, A. C. Allision and A.
Dalgarno, {\em J Chem Phys}, {\bf 81}, 570 (1984).
\bibitem{Da73} A. Dalgarno, J. H. Black and J. C. Weisheit, {\em ApL },
{\bf 14}, 77 (1973).
\bibitem{Ch75} S.-I. Chu and A. Dalgarno, {\em Proc Roy Soc}, {\bf A343},
191 (1975).
\bibitem{Ro82} W. Roberge and A. Dalgarno, {\em ApJ}, {\bf 255}, 176 (1982).
\bibitem{Bu80} S. E. Butler, T. G. Heil and A. Dalgarno, {\em ApJ}, {\bf 241},
442  (1980).
\bibitem{Da90} A. Dalgarno, M. L. Du and J. H. You, {\em ApJ},
{\bf 349}, 675 (1990).
\bibitem{Be08} C. Berteloite, S. D. Le Picard, P. Birza, M.-C. Gazeau, A.
Canosa, Y. Benila, and I. R. Sims, {\em Icarus}, {\bf 194}, 746 (2008).
\bibitem{Mc03} B. J. McCall, A. J. Huneycutt, R. J. Saykally, T. R. Geballe,
N. Djuric, G. H. Dunn, J. Semaniak, O. Novotny, A. Al-Khalill,
A. Ehlerding, F. Hellberg, S. Kalliori, A. Neau, R. Thomas, F. Osterdahl
and M. Larsson, {\em Nature}, {\bf 422}, 500 (2003).
\bibitem{Mc06} M. C. McCarthy, C. A. Gottlieb, H. Gupta and P. Thaddeus,
{\em ApJ}, {\bf 652}, L141 (2006).
\bibitem{Ba06} R. J. Barber, J. Tennyson, G. J. Harris and R. N. Tolchenov,
{\em MNRAS}, {\bf 368}, 1087 (2006).
\bibitem{Ti07} G. Tinetti, A. Vidal-Madjar, M.-C. Liang, J.-P. Beaulieu, Y.
Yuk, S. Carey, R. J. Barber, J. Tennyson, I. Ribas, N. Allard, G. E.
Ballester, D. K. Sing and F. Selsis, {\em Nature}, {\bf 448}, 169 (2007).
\bibitem{Ha08} G. J. Harris, F. C. Larner, J. Tennyson, B. M.
Kaminsky, Ya. V. Pavlenko and H. R. A. Jones, {\em MNRAS}, {\bf 390},
143 (2008).
\bibitem{Ca98a} P. Caselli, T. I. Hasegawa and E. Herbst, {\em ApJ},
{\bf 495}, 309 (1998). 
\bibitem{Gr01} N. J. B. Green, T. Toniazzo, M. J. Pilling, D. P. Ruffle,
N. Bell and T. W. Hartquist, {\em A\&A}, {\bf 375}, 1111 (2001).
\bibitem{Ba07} B. Barzel and O. Biham, {\em J Chem Phys}, {\bf 127},
114703-1 (2007).
\bibitem{Wi07} D. A. Williams, W. A. Brown, S. D. Price, J. M. C. Rawlings
and S. Viti, {\em A\&G}, {\bf 48}, 1.25 (2007).
\bibitem{Ob07} K. I. Oberg, G. W. Fuchs, Z. Awad, H. J. Fraser, S. Schemmer,
E. F. van Dishoeck and H. Linnartz, {\em ApJ}, {\bf 662}, L23 (2007).
\bibitem{Nu06} J. A. Nuth III and F. T. Ferguson, {\em ApJ}, {\bf 649}, 1178
(2006).
\bibitem{Bh07} J. Bhatt and I. J. Ford, {\em MNRAS}, {\bf 382}, 291 (2007).
\end{thebibliography}
\end{document}